\documentclass{article}

\usepackage[active]{srcltx}
\usepackage{latexsym}
\usepackage{amsfonts,amsmath,amssymb}
\usepackage{graphicx}

\def\eq#1{{Eq.~(\ref{#1})}}

\def\cn{{CosMIn}}

\title{\cn: The Solution to the Cosmological Constant Problem}

\author{Hamsa Padmanabhan, T. Padmanabhan\\
IUCAA, Post Bag 4, Ganeshkhind,
 Pune - 411 007, India.\\
email: hamsa@iucaa.ernet.in, paddy@iucaa.ernet.in}

\date{ }

\begin{document}

\maketitle

\begin{abstract}
The current acceleration of the universe can be modelled in terms of a cosmological constant $\Lambda$. We show that the extremely small value of  $\Lambda L_P^2 \approx 3.4\times 10^{-122}$, the holy grail of theoretical physics, can be understood in terms of a new,  dimensionless, conserved number \cn, which counts the number of modes  crossing the Hubble radius during the three phases of  evolution of the universe. Theoretical considerations suggest that $N \approx 4\pi$. \textit{This single postulate leads us to  the correct, observed numerical value of the cosmological constant!} This approach also provides a unified picture of cosmic evolution relating the early inflationary phase to the late accelerating phase.\footnote{Essay selected for Honorable Mention in 
 the Gravity Research Foundation Essay Contest 2013.} {\small{


\textit{PACs:} 95.36.+x 04.60.-m 98.80.Es
}}

\end{abstract}

\maketitle

\textbf{Introduction:}
Our description of the cosmos is very tantalizing! It has three distinct phases of evolution, bearing no apparent relation to each other: An early inflationary phase, driven possibly by a scalar field, a late-time accelerated phase, dominated by dark energy,  and a transient phase in between, dominated by radiation and matter.
 
The first and the last phases 
are approximately de Sitter, with  Hubble radii $H_{\rm inf}^{-1}$ and $H_\Lambda^{-1}$, characterized by two dimensionless ratios $\beta^{-1} \equiv H_{\rm inf} L_P $ and $\Lambda L_P^2$, where $L_P \equiv (G\hbar/c^3)^{1/2} $
is the Planck length. If  inflation took place at GUTs scale ($\sim 10^{15}$ GeV), then  $\beta \approx 3.8\times 10^7$, while observations \cite{cosmoreview1, cosmoreview2} suggest that $\Lambda L_P^2 \approx 3.4\times 10^{-122} \approx 3\times e^{-281}$. 
It is expected that  physics  at, say, the GUTs scale will (eventually) determine  $\beta$. But no fundamental principle has been suggested  to explain the extremely small value of $\Lambda L_P^2$, which is related (directly or indirectly) to the cosmological constant problem.  Understanding this issue \cite{ccreview1, ccreview2} from first principles is considered very important in theoretical physics today.

In this essay, we will describe an approach which tackles this problem (for more details, see \cite{hptp}) and also provides a unified picture of cosmic evolution. 
 We show that $\ln (\Lambda L_P^2)$ is related to a 
 dimensionless number (`Cosmic Mode Index', or \cn, $N_c$) that counts the number of modes within the Hubble volume that cross the Hubble radius  between the end of inflation and the beginning of late-time acceleration. \cn\ \textit{is a characteristic number for our universe} and it is possible to argue \cite{tpdetcc} that the natural value for $N_c$ is about $4\pi$; i.e., $N_c= 4\pi \mu$ with $\mu \sim 1$. 
 This single postulate allows us to determine the numerical value of $\Lambda L_P^2$! We obtain  $\Lambda L_P^2 = C\beta^{-2} \exp(-24 \pi^2 \mu)$, where $C $ depends on $n_\gamma/n_m$, the ratio between the number densities of photons and matter. \textit{This leads to the correct observed value of the cosmological constant} for a GUTs scale inflation and the range of $C$ permitted by cosmological observations. 

\begin{figure}
 \begin{center}
  \includegraphics[scale=0.35]{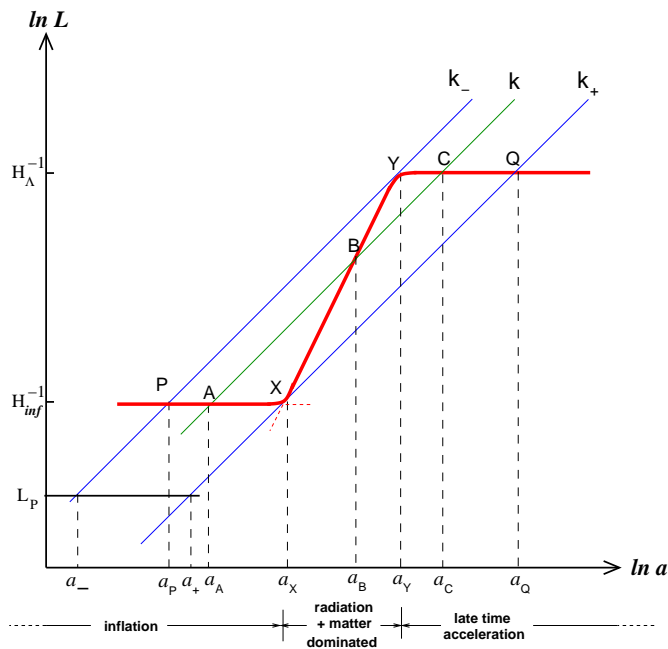}
 \end{center}
\caption{Various length scales in the universe; see text for description.}
\label{fig:hubble}
\end{figure}

\textbf{\cn\ and the cosmological constant:}
A proper length scale  $\lambda_{\rm prop}(a) \equiv a/k$ (labelled by a co-moving wave number, $k$) crosses the Hubble radius  whenever the equation $\lambda_{\rm prop}(a)=H^{-1}(a)$, i.e., $k=a H(a) $ is satisfied. 
For a \textit{generic} mode (see Fig.\ref{fig:hubble}; line marked $ABC$), this equation has three solutions: $a=a_A$ (during the inflationary phase; at $A$), $a=a_B$ (during the radiation/matter dominated phase; at $B$), $a=a_C$ (during the late-time accelerating phase; at $C$).
But modes with $k<k_-$ exit during the inflationary phase and \textit{never} re-enter. Similarly, modes with $k>k_+$ remain inside the Hubble radius and only exit during the late-time acceleration phase. 

The modes with comoving wavenumbers in the range $(k, k+dk)$ where $k=aH(a)$ and $dk=[d(aH)/da] da$ cross the Hubble radius during the interval $(a,a+da)$. The number of modes in a comoving Hubble volume $V_{\rm com}=(4\pi H^{-3}/3a^3)$ with wave numbers in the interval  $(k, k+dk)$ is $dN = V_{\rm com} d^3k/(2\pi)^3 $. Hence, the number of modes that cross the Hubble radius in the interval $(a_1 < a < a_2)$ is given by
\begin{equation}
 N(a_1,a_2) = \int_{a_1}^{a_2} \frac{V_{\rm com} k^2}{2\pi^2} \, \frac{dk}{da}\ da
= \frac{2}{3\pi}\int_{a_1}^{a_2} \frac{d(Ha)}{Ha} 
 = \frac{2}{3\pi}\ln \left(\frac{H_2a_2}{H_1a_1}\right),
\label{Ndef}
\end{equation} 
where we have used $V_{\rm{com}}=4\pi/3H^3a^3$ and $k=Ha$.

All the modes which exit the Hubble radius during $a_A<a<a_X$ enter the Hubble radius during $a_X<a<a_B$ (and again exit during $a_Y<a<a_Q$.)
So the number of modes which do this during $PX$, $XY$ or $YQ$ is a \textit{characteristic, `conserved'  number (``\cn")} for our universe, say $N_c$.
Its importance  is related to the the cosmic parallelogram $PXQY$ (Fig.\ref{fig:hubble}) which arises \textit{only} in a universe having three distinct phases. The epochs $P$ and $Q$, limiting the otherwise semi-eternal de Sitter phases, now  have a special significance \cite{bj, tpgrg,tpbj2}.  Modes which exit the Hubble radius before $a=a_P$ never re-enter. On the other hand, the epoch $a=a_Q$ denotes (approximately) the time when the CMBR temperature falls below the de Sitter temperature \cite{bj, tpgrg,tpbj2}. The special role of $PXQY$ makes the value of  \cn\ significant. As shown in Fig.\ref{fig:hubble}, these modes in $PXQY$ (with $k_-<k<k_+$) cross the Planck length during $a_-<a<a_+$. Based on holographic considerations, it is possible to argue \cite{tpdetcc} that Planck scale physics imposes the condition $N_c=N(a_-,a_+)\approx 4\pi$ at this stage. So, by computing \cn\ for the universe, and equating to $4\pi$, we can determine $\Lambda L_P^2$. 

As a quick check on the paradigm $N_c\approx 4\pi$, let us approximate the intermediate phase of the  universe as purely radiation dominated ($H(a)\propto a^{-2}$) and assume Planck scale inflation ($\beta=1$), thereby eliminating \textit{all} free parameters. The above procedure now \cite{tpdetcc} gives:  
\begin{equation}
 \Lambda L_P^2 = \frac{3}{4} \exp(-24\pi^2 \mu); \qquad \mu \equiv \frac{N_c}{4\pi} \, .
\label{holy1}
\end{equation}
Thus,  $\Lambda L_P^2$ is directly related to \cn\ and, in this simple model, there are \textit{no other adjustable parameters}. \eq{holy1}  leads to the observed value $\Lambda L_P^2 = 3.4\times 10^{-122}$ when $\mu =1.18$, showing we are clearly on the right track! 

The presence of matter  and the fact that  the inflationary scale may not be the Planck scale in our universe ($\beta\neq 1$), surprisingly, make the postulate of $N_c = 4\pi$ work \textit{ better} in the real universe and reproduce the observed value of the cosmological constant. In this case, it is simpler to express $\Lambda$ in terms of $N_c, \ \beta$ and  a  variable $\sigma$  defined through
$
\sigma^4 \equiv(\Omega_R^{3}/\Omega_m^4) [1-\Omega_m - \Omega_R]. 
$
Even though the values for $\Omega_R$ and $\Omega_m$ depend on the epoch  $t=t_*$ at which they are measured, the value of $\sigma$ is the same at all epochs. (It is an example of an \textit{epoch-invariant parameter} and, of course, the value of $\Lambda L_P^2$ can only depend on such epoch-invariant parameters \cite{hptp}). 
Determining $N_c$ in terms of $\Lambda L_P^2, \ \beta$ and  $\sigma$, and expressing  $\Lambda L_P^2$ in terms of the other parameters, we get:
\begin{equation}
 \Lambda L_P^2 = \beta^{-2} C(\sigma)\exp[-24\pi^2\mu];\qquad \mu \equiv \frac{N_c}{4\pi} \,
\label{holy2}
\end{equation} 
where $C(\sigma) = 12 (\sigma r)^4\, (3r+4)^{-2}$ and $r$ satisfies the quartic equation $\sigma^4 r^4 =   (1/2) r + 1 $. 
Given the numerical value of $\sigma$,  the 
inflation scale determined by $\beta$, and our postulate $\mu =1$, we can calculate the value of $\Lambda L_P^2$ from \eq{holy2}.

\begin{figure}
 \begin{center}
  \includegraphics[scale=0.9]{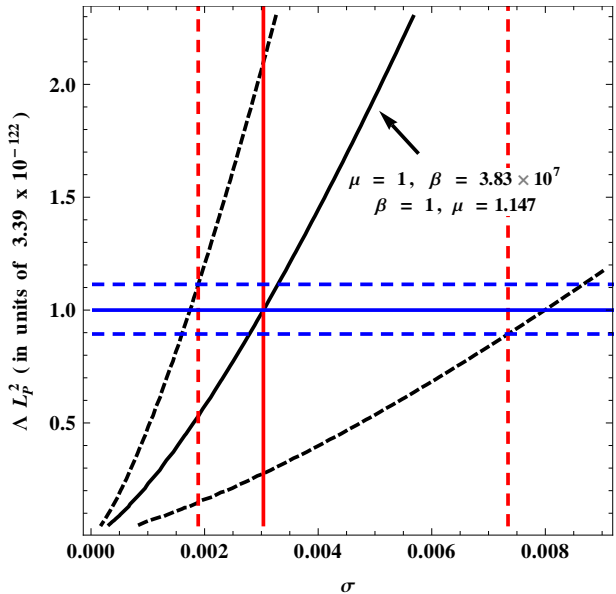}
 \end{center}
\caption{Determination of $\Lambda L_P^2$; see text for discussion.}
\label{fig:lambdasigma}
\end{figure}

 The result in \eq{holy2} is summarized in Fig.\ref{fig:lambdasigma}. 
 The thick black curve is obtained from \eq{holy2} if we take $\mu=1$  and $\beta=3.83 \times 10^7$ (corresponding to the inflationary energy scale of $V_{\rm inf}^{1/4}=1.16 \times 10^{15}$ GeV) and leads to the observed (mean) value of $\Lambda L_P^2=3.39 \times 10^{-122}$ (horizontal unbroken, blue line). Observational constraints \cite{cosmoreview1, cosmoreview2} lead to $\sigma= 0.003^{+ 0.004}_{-0.001}$
  (three vertical, red lines) and $\Lambda L_P^2=(3.03 - 3.77) \times 10^{-122}$ (horizontal, broken blue lines).  This cosmologically allowed range in $\sigma$ and $\Lambda L_P^2$  is bracketed by the two broken black curves obtained by varying $\beta$ in the range $(2.64 - 7.29) \times 10^7$  (i.e, $V_{\rm inf}^{1/4} =  (0.84 - 1.4 )\times 10^{15}$ GeV). So, for an acceptable range of energy scales of inflation, and for the  range of $\sigma$ allowed by cosmological observations, our postulate $N_c = 4\pi$ gives the correct value for the cosmological constant.

 Since our results only depend on the combination $\beta^{-2}\exp(-24\pi^2\mu)$, the same set of curves arise in a Planck scale inflationary model ($\beta=1$) with $\mu$ in the range $(1.144-1.153)$. 
 There are three conceptually attractive features about Planck scale inflation  with $\beta =1$.
First, it eliminates one free parameter, $\beta$, and gives a direct relation between the two scales $\Lambda$ and $L_P^2$ which occur in the Einstein-Hilbert action. (The dependence of the result on $\sigma$ is weak and can be thought of as a matter of detail, like, for example, the fine structure correction to  spectral lines beyond Bohr's model). Second, we can think of the  intermediate phase as a mere transient connecting two de Sitter phases (the chicken is just the egg's way of making another egg!), both of which are semi-eternal. Since the de Sitter universe is time-translationally invariant,  it is a natural candidate to describe the geometry of the universe dominated by a single length scale  --- $L_P$ in the initial phase and $\Lambda^{-1/2}$ in the final phase. The quantum instability of the de Sitter phase at the Planck scale  can lead to cosmogenesis and the transient radiation/matter dominated phase, which gives way, eventually, to the  late-time acceleration phase.  Finally, the argument for $N_c \approx 4\pi$ is quite natural with Planck scale inflation.  The transition at $X$,  entrenched in Planck scale physics in such a model,  can easily account for deviations of $\mu$ from unity.

\textbf{An integrated view of cosmology:}      
 The standard approach to cosmology treats the evolution of the universe in a fragmentary manner, with  Planck scale physics, the inflationary era, the matter sector properties and the late-time acceleration each introducing their own parameters --- like $ L_P^2, \ E_{\rm inf}, \ (n_m/n_\gamma), \ \Lambda $ --- all independently specified, bearing no relation with each other. Even if GUTs scale physics (eventually) determines $E_{\rm inf}$ and $(n_m/n_\gamma)$, there is still no  link between these parameters, $L_P$ and $\Lambda$. 

\textit{In striking contrast, our  paradigm, the postulate $N_c=4\pi$ acts as the connecting thread leading to a unified, holistic approach to cosmic evolution. }
In fact,  when $\sigma \ll 1$, one can write \eq{holy2} (with $\mu=1$) as:
\begin{equation}
L_P^4\rho_\Lambda=K\left(\frac{M_P}{m}\right)^2 \left(\frac{n_\gamma}{n_m}\right)^2 \beta^{-3}\exp(-36\pi^2)
 \label{hep}
\end{equation} 
where  $K\equiv(\pi^{11/2}/\zeta(3)^2)(1/6^3 10^{3/2})\approx 0.055, \
 \rho_m\equiv mn_m$ with $m$ being the (mean) mass of the particle contributing to matter density, and
$M_P$ being the Planck mass. 
In a consistent quantum theory of gravity, we expect
 inflation (which determines $\beta$) and genesis of matter (which determines $m$ and $n_m/n_\gamma$) to be related to Planck scale physics such that our fundamental relation in \eq{hep} holds. 

Solving the cosmological constant problem \textit{by actually determining its numerical value} has \textit{not} been attempted before. 
This approach is similar in spirit to the Bohr model of the hydrogen atom, which used the  postulate $J=n\hbar$ to explain the  hydrogen spectrum. Here, our postulate $N_c=4\pi$, captures the essence and explains the value of $\Lambda L_P^2$.
This is simpler and more elegant than many other ad-hoc assumptions made in the literature \cite{ccreview1, ccreview2} to solve the cosmological constant problem. \textit{More importantly, we do know that this postulate is correct!} The value of \cn\ 
can be determined directly from the observed value of $\Lambda$ \textit{as well as} other cosmological parameters. 
We would then find that it is indeed very close to $4\pi$. 

\textbf{Why does it work?} Recent work \cite{tpdetcc} has shown that cosmic evolution can be thought of as a quest for \textit{holographic equilibrium}. One can associate with the \textit{proper} Hubble volume $V_{prop}\equiv 4\pi/3H^3$, the numbers, 
\begin{equation}
N_{sur}\equiv 4\pi H^{-2}/L_P^2;\qquad N_{bulk}\equiv-\epsilon E/(1/2)k_BT 
\end{equation}
which count the surface and bulk degrees of freedom, where $E=(\rho+3p)V_{prop}$ is the Komar energy, $T=H/2\pi$ is the analogue of the horizon temperature and $\epsilon=\pm1$ is chosen to keep $N_{bulk}$ positive. Clearly,  $|E|=(1/2)N_{bulk}k_BT$ denotes equipartition of energy.
Holographic equipartition is the demand that $N_{sur}=N_{bulk}$, which holds in a de Sitter universe with $p=-\rho, \ \epsilon=1, \ H^2=(8\pi/3)\rho$. When the universe is not pure de Sitter, we expect the holographic discrepancy between $N_{sur}$ and $N_{bulk}$ to drive the expansion of the universe, which suggests \cite{tpbj2} the law:
\begin{equation}
\frac{dV_{prop}}{dt}=L_P^2(N_{sur}-\epsilon N_{bulk})
\end{equation} 
\textit{Incredibly, this leads to the standard Friedmann equation for cosmic expansion, but now obtained} without \textit{using the field equations of general relativity!} The right hand side is (nearly) zero in the initial and final phases (with $V_{prop} \approx$ constant) and cosmic expansion in the transient phase can be interpreted as a quest towards holographic equipartition. It is then natural to associate a number $N_{sur}=4\pi L_P^2/L_P^2=4\pi$ with the modes, when they cross the Planck scale. 

Clearly, such a quantum gravitational imprint has far reaching consequences, culminating in the solution to the cosmological constant problem itself.

\textbf{Acknowledgements:}
T.P's research is partially supported by the J. C. Bose research grant 
   of DST, India. H.P's research is supported by the SPM research grant of CSIR, India.
   We thank Sunu Engineer for useful comments.


\begin{thebibliography}{000}


\bibitem{cosmoreview1}  O. Lahav and A. R. Liddle (2010),  in ``Review of Particle Physics", K. Nakamura et al. (Particle Data Group) , J. Phys. G: Nucl. Part. Phys. \textbf{37},  075021

\bibitem{cosmoreview2} Hinshaw, G., et al. (2012),   [arXiv:1212.5226]


\bibitem{ccreview1} For a recent review, see J. Martin (2012),  C. R. Physique \textbf{13}, 566 [arXiv:1205.3365]

\bibitem{ccreview2} For a classification of approaches to the cosmological constant problem, see
S. Nobbenhuis (2006), Found. Phys. \textbf{36}, 613 [arXiv:gr-qc/0411093].

\bibitem{hptp} T. Padmanabhan, Hamsa Padmanabhan (2013), paper in preparation

\bibitem{bj} J. D. Bjorken (2004),  [arXiv:astro-ph/0404233]

\bibitem{tpgrg} T. Padmanabhan (2008), Gen.Rel.Grav. \textbf{40}, 529 [arXiv:0705.2533]

\bibitem{tpbj2} 
T. Padmanabhan (2012),  Res. Astro. Astrophys. \textbf{12}, 891 [arXiv:1207.0505]


\bibitem{tpdetcc} T. Padmanabhan (2012) [arXiv:1210.4174]





 \end{thebibliography}
\end{document}